\documentclass[12pt]{article}
\begin{document}

\title{A Non-Abelian Fourier Transform for Gauge Theories}

\author{Grigorii B. Pivovarov$^{1,2}$ and James P. Vary$^{1}$\\
$^1$\small\it Department of Physics and Astronomy, 
Iowa State University, Ames, Iowa 50011, USA\\
$^2$\small\it Institute for Nuclear Research, Moscow 117312, Russia}

\date{November 28, 2001}
\maketitle

\abstract{
We consider $SU(2)$ gauge potentials over a space with a compactified 
dimension.
A non-Abelian Fourier transform of the gauge potential in the compactified
dimension is defined in such a way that the Fourier coefficients are
(almost) gauge invariant. The functional measure and the gauge field 
strengths are expressed in terms of these Fourier coefficients.
The emerging formulation of the non-Abelian gauge theory turns out to be
an Abelian gauge theory of a set of fields defined over the 
initial space with the compactified dimension excluded. The Abelian 
theory contains an Abelian gauge field, a scalar field,  and an 
infinite tower of vector matter fields, some of which carry Abelian charges. 
Possible applications of this 
formalism are discussed briefly.}

\section{Introduction}

Despite the years of efforts, non-Abelian gauge theories have resisted 
the attempts to 
obtain a mathematically rigorous, or operationally efficient definition
beyond perturbation theory. 
A particular way to illustrate this assessment is to check what is 
presently known beyond
perturbation theory about the ultraviolet renormalization of gluodynamics
(see \cite{Shuryak,Lattice} for recent studies).
Simply stated, to date we have only a reasonable expectation
that the theory exhibits asymptotic freedom at weak coupling.

This state of affairs justifies further efforts to reformulate 
the theory, and 
reveal nontrivial observables that may be computable beyond perturbation 
theory. By nontrivial observables we mean the observables that exhibit 
nontrivial ultraviolet behavior 
characteristic of the dynamics. 

Candidates for this role can be found in \cite{KMPV}. In this paper,
the light-front formulation of the non-Abelian gauge theories has been
reconsidered. A conclusion has been made that there are observables
whose dynamics can be split into sectors, each sector representing
quantum mechanics of a finite number of degrees of freedom. If this 
conclusion is true, we indeed have found 
the desired observables,
since many practical methods can be used to solve quantum mechanics
with a finite number of degrees of freedom.
 
However, the conclusion of \cite{KMPV} is not in agreement with some 
other 
results (for a discussion and references, see
\cite{KMPV}). The competing view holds that the light-front formulation is 
plagued with
complications related to the treatment of the zero modes.

The formulation of \cite{KMPV} is tied to the Hamiltonian framework. While
suitable for setting the stage, it is not necessarily the most convenient for
implementing  the expected symmetries.

In this paper, we avoid reference to the light-front formulation and formulate the non-Abelian Fourier transform of the 
gauge potentials
implicitly present in \cite{KMPV}.
The key feature of the emerging Fourier coefficients of the gauge potentials
is that they are 
almost gauge invariant. 

The source of the residual gauge non-invariance of the Fourier 
coefficients is a compactification of the direction along which we consider the
Fourier transform. We compactify along this direction because we want to
have a discrete Fourier series. In this case, when a direction 
of the space-time is compactified, there are gauge transformations
that leave the component of the gauge field along the compactified direction  
intact. Namely those transformations act nontrivially on the Fourier
coefficients we define.

In the next section, we consider a non-Abelian gauge field (to be specific, 
an $SU(2)$ gauge field $A_\mu$) over a manifold compactified in a 
selected direction. 
Using the known
facts about the covariant derivative in the compactified direction, and 
the Wilson loop embracing the compactified direction, we determine a 
set of almost gauge invariant Fourier components of the gauge potential.
The standard formal functional measure
$\prod_{\mu,z} dA_\mu(z)$ is expressed in terms of the Fourier coefficients.

In the third section, we express the field strengths
$F_{\mu\nu}$ in terms of the new variables.
With these tools, we can express the entire gauge theory in 
terms of the new variables. The emerging formulation of the non-Abelian gauge
theory turns out to be an Abelian gauge theory of fields defined over the
initial space-time with the compactified direction excluded.

The field content of the new formulation is as follows: There is an Abelian 
gauge field, a scalar field (the trace of the Wilson loop embracing the 
compactified direction), and an infinite tower of vector matter fields
some of which are carrying Abelian charges.

Here we mention that resolving an Abelian theory inside a non-Abelian theory
is an important ingredient in a confinement scenario
\cite{thooft} (see also \cite{shabanov1} for a recent discussion).

In the last section, we conclude with a discussion of the prospects for employing
the non-Abelian Fourier transform with the light-front formulation, and, in 
particular, for calculations of the 
nonperturbative  ultraviolet 
renormalization of gluodynamics. As our formulation is gauge invariant, we
are able to formulate a gauge invariant ultraviolet regularization in a manner 
alternative to \cite{Bardeen}.

\section{The Fourier Data and the Measure}

Consider a space $S^1\times M$ and the gauge potential over it, $A_{\mu}=
(A,A_i)$.
Let the coordinate along the circle $S^1$ be $x$, and the manifold $M$ has 
$y^i$ as its
coordinates. Let also the component $A(x,y)$ define the parallel 
transport along the circle, and the components of the potential $A_i(x,y)$   
 define the parallel transport along the curves in $M$. 

To be specific, we consider only the case of $SU(2)$ gauge fields. We do 
not expect any complication in generalizing  to $SU(N)$.  
So, $A_\mu$ are traceless, 
Hermitian, $2\times2$ matrix functions periodic in $x$.

Consider the space of matrix-valued functions with the scalar product
\begin{equation}
\label{scalarprod}
\langle\Phi|\Psi\rangle = \int\,dx\,2 \,{\rm Tr}\, 
\big[\Phi^\dagger\Psi\big].
\end{equation}
The dagger signifies Hermitian conjugation and the normalizing factor $2$
makes the Pauli matrixes divided by two orthonormal at a unit length 
of the circle.
Our functions will depend on the location $y^i$ on $M$, but 
this is considered as a parametric dependence of the functions defined 
over the circle.  

With respect to this scalar product, the operator $iD$ of the covariant 
derivative along the $x$-direction is Hermitian. Explicitly, 
$iD\Phi=i\partial_x\Phi + g[A,\Phi]$. The characteristic properties of 
the spectrum of $iD$ for the circle of length $L$ are as follows: 
\begin{itemize}
\item{It is a spectrum of 
the conventional derivative
on a circle with each level split into three levels (the case of $SU(2)$)}

\item{One of the levels of each triple stays as if it were a 
conventional derivative, i.e., it is $2\pi n/L$ with integer $n$}

\item{Another level of the triple goes above the value $2\pi n/L$ by the 
increment $\phi/L$}

\item{The last level of the triple goes down by the decrement $\phi/L$}

\end{itemize}
The splitting parameter $\phi$ never goes beyond $2\pi$. It is related to the 
value of the Wilson loop 
$w\equiv {\rm Tr}\big[{\rm P} \exp(ig\int dx A)\big]/2$ 
(${\rm P}$ denotes the path ordering)
in the following way: $\phi = 2\, arccos(w)$ 
(we assume that the $arccos$ takes 
the values in the range from $0$ to $\pi$).
Note that the levels forming the triple above are not 
always selected from the spectrum by the condition that they are
close to one another. \footnote{This is the case though for the positive values of $w$
(here $w$ ranges from $-1$ to $+1$).} When $w$ approaches zero from the positive
side, the upper level of the 
triple associated with the level $2\pi n/L$ collides
with the lower level of the triple associated with the 
level $2\pi (n+1)/L$. As $w$ further decreases, the upper level of the
triple associated with $2\pi n/L$ approaches the value $2\pi (n+1)/L$.

The above facts can be checked using explicit formulas of \cite{KMPV}. 
Evidently, the spectrum of $iD$ is gauge invariant (when the
potential $A$ undergoes a gauge transformation, the spectrum of $iD$ 
remains intact). We denote the eigenvalues 
$p(n,\sigma)=2\pi n/L +\sigma\phi/L$, or just $p$ for brevity, 
$\sigma=(-1,0,+1)$.

An orthonormal set of eigenfunctions $\chi_{p(n,\sigma)}$  can be 
chosen in such a way that its members transform 
uniformly under the gauge transformations. This follows from the fact that a 
covariant derivative of a uniformly transforming object transforms in 
the same uniform way. Therefore, if $\chi_p$ is a normalized eigenfunction of
$iD$ at a certain gauge, $U\chi_pU^\dagger$ is a normalized 
eigenfunction of the covariant derivative involving the gauge transformed
field $U(A-\partial_x/(ig))U^\dagger$.

We use the above set of normalized eigenfunctions $\chi_p$ to construct
gauge invariants. For example, 
$F_{\mu\nu}^p\equiv\langle\chi_p|F_{\mu\nu}\rangle$ is a gauge
invariant, because the scalar product (\ref{scalarprod}) involves the trace, 
and both $\chi_p$ and the field strength tensor $F_{\mu\nu}$ are 
transformed uniformly. The above notation with the superscript $p$ 
for the projection of a field onto
the eigenvector $\chi_p$ will be in regular use throughout the text.

We first consider the component $A$ of the gauge field along the circle 
as given, and use it to construct gauge invariants containing 
all the information about the rest of the gauge field components $A_i$.

We decompose the fields $A_i$:
\begin{equation}
\label{decomposition}
A_i=A_i^\perp+A_i^\|,\,A_i^\perp\equiv\sum_{p\neq 0}\chi_pA_i^p,\,
A_i^\|\equiv\chi_0A_i^0.
\end{equation}
In words, we separate the zero mode component (the component along the 
eigenvector of $iD$ with zero eigenvalue) from the rest of the components.
We will treat $A_i^\perp$ and $A_i^\|$ differently. 

For treating $A_i^\perp$, we construct from $A$ and $\chi_p$ an 
object living in the
subspace transverse to the zero mode and transforming under the gauge 
transformations exactly in the way the field $A_i^\perp$ does. Indeed, 
let us check the gauge transformation of
\begin{equation}
\label{B}
B_i\equiv\sum_{p\neq0}\frac{i}{p}\chi_p\langle\chi_p|\partial_iA\rangle.
\end{equation}
Consider the infinitesimal gauge transformation $\delta_\alpha A=D\alpha$.
It generates the following transformation of the coefficients 
$(\partial_i A)^p$ of the expansion (\ref{B}):
\begin{equation}
\label{trasform}
\delta_\alpha\langle\chi_p|\partial_i A\rangle=
\langle-ig[\chi_p,\alpha]|\partial_i A\rangle +
\langle\chi_p|-ig[\partial_iA,\alpha]\rangle +
\langle\chi_p|D\partial_i\alpha\rangle.
\end{equation}
Here, the first term in the rhs is the transformation of $\chi_p$, 
and the last two terms correspond to the two ways of placing the 
partial derivative in $\partial_iD\alpha$. Clearly, the first two terms in
the rhs cancel against each other, and $D$ in the last term 
can be replaced by the factor $p/i$ matching the factor $i/p$ in (\ref{B}).

The net result for the gauge transformation of the components $B^p_i$
is as follows:
\begin{equation}
\label{Btransf}
\delta_\alpha B_i^p = \langle\chi_p|\partial_i\alpha\rangle.
\end{equation}
This is exactly how the components of the gauge field are transformed:
\begin{equation}
\label{Atransf}
\delta_\alpha A_i^p = \langle\chi_p|\partial_i\alpha\rangle.
\end{equation}

We conclude that
\begin{equation}
\label{Psi}
\Psi_i^p\equiv A_i^p - B_i^p 
\end{equation}
are gauge invariant. These are the variables we suggest as part of our
compete set of new variables. As they are obtained by a shift from 
the variables $A_i^p$, there is no Jacobian involved in changing over 
from $A^p_i$-variables to the $\Psi_i^p$-variables
in the functional integral\footnote{There is a subtlety here: we imply that
$\chi_p$ is independent of $A_i$ and is determined by $A$. It 
contradicts the requirement that $\chi_p$ are transformed uniformly: they 
do not change
if $A$ does not change. Therefore, our new variables are gauge 
invariant
only with respect to the gauge transformations that transform $A$ 
nontrivially. The left-over gauge transformations
of the new variables show up as an Abelian gauge
symmetry of the theory formulated in terms of the new variables.}.

Now let us treat $A_i^\|$ (see (\ref{decomposition})). 
Namely, consider the zero modes of $A_i$, $A_i^0=\langle\chi_0|A_i\rangle$. 
We want to find gauge invariants that would be in one-to-one 
correspondence with $A_i^0$. To do it,  let us form a new object out of 
$B_i$ (see (\ref{B}) for the definition) and $A_i^0$:
\begin{equation}
\label{C}
C_i\equiv B_i +\chi_0 A_i^0.
\end{equation}

The point here is that $C_i$ transforms under the gauge transformations 
exactly as a gauge field:
\begin{equation}
\label{Ctransf}
\delta_\alpha C_i = D_i[C]\alpha,
\end{equation}
where $D_i[C]\alpha=\partial_i\alpha-ig[C_i,\alpha]$. It follows from the 
fact that the non-zero modes of $C_i$ coincide with the non-zero modes of
$B_i$, whose transformation we have studied above. Therefore, for any $p$
(including $p=0$) the transformation of the components of
$C_i$ is as it should be for a gauge field: 
$\delta_\alpha C_i^p = \langle\chi_p|\partial_i\alpha\rangle$, 
which implies (\ref{Ctransf}).

Using this $C_i$, we define another gauge invariant:
\begin{equation}
\label{G}
G_i \equiv 
\frac{\sqrt{L}\sigma_p }{g}\langle\chi_p|iD_i[C]\chi_p\rangle.
\end{equation}
The fact that it is a gauge invariant is implied by the transformation 
properties of the field $C_i$ mentioned above. The factor in the rhs
involving the sign $\sigma_p$ is chosen to 
simplify $G_i$; $\sigma_p$ in that factor is determined by $p$:
$p=2\pi n_p/L+\sigma_p\phi/L$.
Naturally, (\ref{G}) is valid only for $\sigma_p\neq 0$. 
As we will 
see, $G_i$ defined in this way is independent of $p$ as soon as 
$\sigma_p\neq 0$.

The key property of $G_i$ is that it is uniquely related to the zero mode 
$A_i^0$. To see this, we substitute the definition of $C_i$ into the 
definition of $G_i$, and obtain:
\begin{equation}
\label{Gexplicit}
G_i=\frac{\sqrt{L}\sigma_p }{g}
\Big(\langle\chi_p|i\partial_i\chi_p\rangle + 
A_i^0\langle\chi_p|g[\chi_0,\chi_p]\rangle\Big).
\end{equation}
The part linear in $B_i$ does not contribute because it is a sum of terms 
proportional 
to $\langle\chi_p|[\chi_{p'},\chi_p]\rangle$ at $p'\neq0$, which is zero. 
This is the case because the above commutator, if nonzero, is an 
eigenfunction of $iD$ with the eigenvalue $p'+ p$, which is necessarily 
orthogonal to $\chi_p$.

We further clarify what is $G_i$ by computing the commutator in the
rhs of (\ref{Gexplicit}):
\begin{equation}
\label{commutator}
[\chi_0,\chi_p] = \frac{\sigma_p }{\sqrt{L}}\chi_p.
\end{equation}
The easiest way to obtain this expression is to go over to the gauge
in which $A$ is diagonal and independent of $x$. In this gauge,
\begin{eqnarray}
\label{lcchi}
\chi_{p(n,0)}& =& \frac{\exp\Big(-i\frac{2\pi n}{L}x\Big)\sigma^3}{2\sqrt{L}},
\nonumber\\
\chi_{p(n,\pm)}& =& 
\frac{\exp\Big(-i\frac{2\pi n}{L}x\Big)(\sigma^1\pm
i\sigma^2)}{2\sqrt{2L}}.
\end{eqnarray}
It can be explicitly checked that these are indeed the desired eigenfunctions
of $iD$ if $A$ is diagonal and independent of $x$. And (\ref{commutator})
holds in this gauge. It remains to notice that the relation (\ref{commutator})
is gauge invariant to conclude that (\ref{commutator}) is valid in any gauge.
    
Plugging (\ref{commutator}) into (\ref{Gexplicit}) we obtain that
\begin{equation}
\label{GevenMoreExplicit}
G_i=\frac{\sqrt{L}\sigma_p }{g}
\langle\chi_p|i\partial_i\chi_p\rangle + A_i^0.
\end{equation}
The rhs of the above equation formally depends on $p$, but in fact it does not.
To see this, again go over to the gauge where the eigenfunctions $\chi_p$ are 
given by (\ref{lcchi}). In this gauge, the first term in the rhs of 
(\ref{GevenMoreExplicit}) disappears, because the eigenfunctions (\ref{lcchi})
are independent of $y^i$. Since $G_i$ is gauge invariant, checking that it is 
independent of $p$ in any gauge suffices to conclude that it is indeed 
independent of $p$. 

We conclude that $G_i$ is, on the one hand, gauge invariant, and, on the 
other hand, coincides with $A_i^0$ up to a shift defined in terms of $\chi_p$.

Ultimately, the set of the Fourier data obtained by
applying the Fourier transform over the circle to $A$, $A_i$ are $w$ 
(the Wilson loop along the compactified direction), $\Psi_i^p$, and
$G_i$ (see Eqs. (\ref{Psi}) and (\ref{G}) for the definitions).
We suggest to use the Fourier data as a new set of the field variables.
The new field variables constitute an infinite set of fields over $M$.

At this point we mention that, in analogy 
with the ordinary Fourier transform of a real function, the Fourier components
$\Psi^p_i$ of the Hermitian matrix field 
$\Psi_i\equiv A_i^\perp-B_i$ satisfy the
condition of complex conjugation $\Psi_i^{-p}=(\Psi_i^p)^\dagger$, where 
the dagger means complex conjugation. 

Next, we derive the measure of the functional integral in terms of the
new variables with the following chain of variable transformations:
\begin{equation}
\label{chain}
\{A,A_i\}\rightarrow \{A,\Psi_i^p,G_i\}\rightarrow \{w,\Psi^p_i,G_i\}.
\end{equation}
The first transformation does not introduce any Jacobian, since it is  
a rotation and a shift  of an orthonormal base in the functional space
(see Eqs. (\ref{Psi}) and (\ref{G})).
The last transformation involves only $A$, and it is known how to go over from 
integration of a gauge invariant function over $A$ to the integration
of the same function over the Wilson loop $w$ (see, for example, 
\cite{Shabanov}). Intuitively, to go over from the integration over
$A$ to the integration over $w$, we should replace the set of variables $A$ 
with the parameters of the gauge transformations $\alpha$, 
and with the gauge invariant $w$. It may seem that $\alpha$ contains
the same number of parameters as $A$ does, and we don't need
 $w$. That would be the case if there would be no gauge 
transformations that leave $A$ invariant. In fact, the 
nontrivial transformations of $A$ can be singled out by the requirement
$\langle\chi_0|\alpha\rangle = 0$, because the gauge transformation
of $A$ generated by $\chi_0 $ is trivial ($D\chi_0=0$). So, there is a 
correspondence between the variables encoded in $A$, and the variables
defining a vector of the space of the gauge transformations
transverse to $\chi_0$. The correspondence is given by the relation
$\delta_\alpha A=D\alpha$. Therefore, the transformation of the variables 
from $A$ to 
the variables parameterizing the gauge group generates the Jacobian
$|{\rm det}[D']|$, where the prime denotes that the zero eigenvalue of $D$
should be excluded from the computation of the determinant. It is known 
how to compute this determinant, see \cite{Shabanov}.
For $SU(2)$, $|{\rm det}[D']| = 1-w^2$.
The gauge invariant content of $A$ is encoded in $w$, and the Jacobian of the
transformation from the ``gauge invariant piece of $A$'' to $w$ can 
be computed in any convenient gauge, in particular, in the gauge where the
eigenfunctions $\chi_p$ are given by (\ref{lcchi}). This extra Jacobian is
$1/\sqrt{1-w^2}$. 
Therefore, the correct replacement of the measure of 
integration over $A$ is as follows:
\begin{equation}
\label{Ameasure}
\prod_{x,y}dA(x,y)\rightarrow \prod_y \sqrt{1-w^2(y)}dw(y).
\end{equation}
We stress that this substitution is valid in the integrals of gauge 
invariant functions. Another point to make is that the integration over 
each $w(y)$ runs from $-1$ to $1$.

We summarize the above by stating that the partition function of the 
gluodynamics over $S^1\times M$ is
\begin{eqnarray}
\label{partition}
Z&=&\int
\prod _{p>0}\prod _{y\in M,i}d(\Psi_i^p(y))^\dagger d\Psi_i^p(y)  
\prod _{y\in M,i}dG_i(y)\times\nonumber\\
&&\times\prod _{y\in M}dw(y)\sqrt{1-w^2(y)}
\times\exp\big(iS_{glue}\big).
\end{eqnarray}
Here we used the fact that $\Psi^{-p}_i=(\Psi_i^p(y))^\dagger$ to restrict 
the product over $p$ to the positive values (recall that 
$p=(2\pi n+\sigma\phi)/L$).

To complete the picture, we have to express $S_{glue}$ in terms of the 
new variables. As $S_{glue}$ is expressible in terms of the field strength
tensor $F_{\mu\nu}$, it suffices to express 
$F^p_{\mu\nu}=\langle\chi_p|F_{\mu\nu}\rangle$ in terms of the new 
variables. This will be done in the next section.

\section{Field Strengths in Terms of the New Variables} 

To express the field strengths in terms of the gauge invariant Fourier modes, 
we first recall how $A_i$ is related to $\Psi_i^p$: 
\begin{equation}
\label{A}
A_i=\Psi_i+C_i,
\end{equation}
where $\Psi_i\equiv \sum_{p\neq 0}\chi_p\Psi_i^p$, and $C_i$ is defined in
(\ref{C}).

Expressing $F_{\mu\nu}$ for the case when one of the indexes corresponds to
the compactified direction is simpler than for  $F_{ij}$. We start from
this case. Consider $E_i\equiv F_{xi}$, where the index $x$ corresponds to 
the compactified direction. In terms of the covariant derivative $D$ and 
the gauge potential, $E_i$ reads
\begin{equation}
\label{E}
E_i = DA_i - \partial_i A.
\end{equation}
Substitute into this formula (\ref{A}) and  (\ref{C}) to obtain
\begin{equation}
\label{E1}
E_i = D\Psi_i + DB_i - \partial_i A.
\end{equation}
This is the case because the difference between $C_i$ and $B_i$ is
proportional to $\chi_0$, and, therefore, it is nullified by the action of
$D$. 

Now consider the projection of $E_i$ onto $\chi_p$:
\begin{equation}
\label{Ep}
E_i^p=\frac{p}{i}\Psi^p_i +\frac{p}{i}B_i^p - (\partial_i A)^p.
\end{equation}
Check the definition (\ref{B}) and observe that the second term in 
the rhs above cancels against the third term if $p\neq 0$. Therefore,
$E_i^p=p\Psi_i^p/i$ if $p\neq 0$. If $p =0$, the first two terms do not 
contribute, and, therefore, $E_i^0=-(\partial_i A)^0$. This is a gauge 
invariant:
\begin{center}
$ 
\delta_\alpha (\partial_i A)^0=
\langle-ig[\chi_0,\alpha]|\partial_i A\rangle + 
\langle\chi_0|-ig[\partial_i A,\alpha]\rangle +
\langle\chi_0|D\partial_i \alpha\rangle =0.
$
\end{center}
\noindent The last equality holds because in the last term  
$D$ nullifies $\chi_0$, and the first two terms are canceled against 
each other.

To compute $(\partial_i A)^0$ go over to the gauge where the 
eigenfunctions are given by (\ref{lcchi}) and obtain 
$(\partial_i A)^0=-2\partial_i w/(g\sqrt{L(1-w^2)})$. 

Ultimately, for the Fourier modes $E_i^p$ we have
\begin{eqnarray}
\label{Epexplicit}
E_i^0&=&\frac{2\partial_i w}{g\sqrt{L(1-w^2)}},\\
E_i^{p\neq 0}&=&\frac{p}{i}\Psi_i^p.
\end{eqnarray}

Next step is to treat $F_{ij}$. Use (\ref{A}) and expand $F_{ij}$ 
in powers of $\Psi_i$:
\begin{eqnarray}
\label{power}
F_{ij}& =& 
\Big(\partial_i C_j -\partial_j C_i -ig[C_i, C_j]\Big)\nonumber\\
&+&
\Big(D_i[C]\Psi_j - D_j[C]\Psi_i \Big)- \Big(ig [\Psi_i,\Psi_j]\Big).
\end{eqnarray}
Since $\Psi$ and $F_{ij}$ transform uniformly under gauge transformations,
each bracket in the rhs above transforms uniformly, and the Fourier modes of
each bracket are separately gauge invariant. Therefore, to compute them, we
can use any convenient gauge. 

Start with the first bracket. In the gauge 
where the eigenfunctions are given by (\ref{lcchi}), $C_i=G_i\chi_0$,
and $\chi_0$ is independent of $y^i$. Therefore, in this gauge,
the first bracket becomes $\big(\chi_0(\partial_iG_j-\partial_jG_i)\big)$.
We see that only the zero mode of the first bracket is nonzero, and it equals
$\partial_iG_j-\partial_jG_i$. It can be interpreted as the field strength 
of the Abelian gauge field $G_i$ over the space $M$. 

Consider the second bracket of the rhs of (\ref{power}). Do the same trick: 
go over to the gauge where the eigenfunctions are given by (\ref{lcchi}) and
compute Fourier modes $(D_i(C)\Psi_j)^p$ (for this bracket, both terms 
of the bracket have gauge invariant Fourier modes). In the calculation,
use (\ref{commutator}). The answer is
\begin{center} 
$
(D_i(C)\Psi_j)^p=\partial_i\Psi_j^p-
i\frac{g\sigma_p }{\sqrt{L}}G_i\Psi_j^p
$,
\end{center}
\noindent which looks like a covariant derivative ${\cal D}_i[G]$ of 
an Abelian
theory acting on a vector matter field $\Psi_j^p$ defined over $M$. 
In this, $G_i$ is 
the Abelian gauge field, and the Abelian charge of the field $\Psi_j^p$
is $g\sigma_p /\sqrt{L}$. We note here that the Abelian charge of
fields $\Psi_i^p$ vanishes for the fields with $\sigma_p = 0$ (these
fields are the Fourier modes corresponding to the middle levels
of the triples of the spectrum of $iD$).

Consider now the last bracket of the rhs of (\ref{power}). Expand the 
$\Psi$ fields involved in this bracket into Fourier modes, and use the
above convenient gauge to compute the emerging commutators. The answer for
them generalizes (\ref{commutator}):
\begin{equation}
\label{GeneralCommutator}
[\chi_{p_1},\chi_{p_2}] = 
\frac{s(\sigma_{p_2}-\sigma_{p_1}) }{\sqrt{L}}\chi_{p_1+p_2}.
\end{equation}  
Here, $s(\sigma_{p_2}-\sigma_{p_1})$ is the sign of the difference of the 
signs
if the latter is nonzero, and it is zero if the difference of the signs is 
zero. With this expression, the Fourier modes
of the last bracket of the rhs of (\ref{power})
take the following form:
\begin{equation}
\label{LastBracket}
\big(ig[\Psi_i,\Psi_j]\big)^p=ig\sum_{p_1}
\frac{s(\sigma_{p-p_1}-\sigma_{p_1}) }{\sqrt{L}}
\Psi_i^{p_1}\Psi_j^{p-p_1}.
\end{equation} 

Gathering together the above information 
about the separate brackets of the rhs of
(\ref{power}), we obtain the ultimate expressions for the Fourier modes
of $F_{ij}$:
\begin{eqnarray}
\label{Fij0}
F_{ij}^0&=&{\cal F}_{ij} - 
ig\sum_p
\frac{s(\sigma_{-p}-\sigma_{p}) }{\sqrt{L}}
\Psi_i^{p}\Psi_j^{-p},\\
\label{Fijp}
F_{ij}^{p\neq 0}&=&{\cal D}_i\Psi_j^p-{\cal D}_j\Psi_i^p-\nonumber\\
&&-ig\sum_{p_1}
\frac{s(\sigma_{p-p_1}-\sigma_{p_1}) }{\sqrt{L}}
\Psi_i^{p_1}\Psi_j^{p-p_1}.
\end{eqnarray}
Here, the Abelian strength tensor ${\cal F}_{ij}$ and the Abelian
covariant derivatives ${\cal D}_i\Psi_j^p$ are defined as follows:
\begin{eqnarray}
\label{AbelianF}
{\cal F}_{ij}&\equiv& \partial_i G_j -\partial_j G_i,\\
\label{AbelianD}
{\cal D}_i\Psi_j^p&\equiv&\partial_i\Psi_j^p -ig^p G_i\Psi_j^p.
\end{eqnarray}
Above, the Abelian charge $g^p$ is
\begin{equation}
g^p\equiv \frac{g\sigma_p }{\sqrt{L}}.
\end{equation}

This completes the description of the Fourier components of the field
strength tensor. Let us list the general properties of $SU(2)$ gluodynamics in 
the new variables:
\begin{itemize}
\item{It is an Abelian gauge field theory over $M$}
\item{Apart from the Abelian gauge field, the model contains an infinite
tower of vector matter fields, and a single neutral scalar field 
(the Wilson loop)
taking values in the interval $(-1,1)$}
\item{Generally, there are cubic and quartic interactions between
the matter fields}
\item{There are charged and uncharged vector matter fields}
\end{itemize}

In the next section, we discuss the prospects of
using the new field variables. In particular, we consider 
the case when the compactified direction 
is light-like, and the overall space $S^1\times M$ carries a Minkowski metric.
We do this case because there is a hope based on \cite{KMPV} that not only the 
infinite tower of the matter fields present in the model, but even the 
number of excitations of separate fields can be consistently bounded from
above by specifying a value of the total momentum of the system in the
compactified light-like direction. 

\section{Discussion}

The action and the partition
function of gluodynamics in terms of the new variables 
depend on the relation between the metric on the initial space 
$S^1\times M$ and the compactified direction. 

Two particularly interesting cases are when the metric is Euclidean,
and when the metric is Minkowskian and the compactified
direction is light-like. The former is interesting because it treats the 
case of finite temperature (inverse radius of the circle). The latter is 
interesting because there is a hope based on \cite{KMPV} that restricting
the value of the total momentum of the system along the compactified
direction is a condition stringent enough to leave only a finite number of 
degrees of
freedom. In that case, we would have a unique possibility to study 
the ultraviolet divergences of gluodynamics nonperturbatively
in a framework of quantum mechanics of a finite number of degrees of freedom.
Here we present only initial steps on realizing the second possibility with 
the Minkowskian metric.

Let the initial space $S^1\times M$ be equipped with Minkowski metric 
$g_{\mu\nu}$ with a single time-like direction. 
Let the direction along the circle be light-like.
Let $M=R\times T$, where $R$ is a light-like line, and $T$ is the 
transverse space (for the four-dimensional space-time, $T$ is 
a two-dimensional Euclidean space).
In this section, the coordinate along the circle is denoted as $x^-$.
Let $x^+$ be the coordinate on $R$, and
$y^I$ be the coordinates on $T$. Let the nonzero components of the metric
in the coordinates $x^-,x^+,y^I$ be defined by $g_{+-}=g_{-+}=1$ and
$g_{IJ}=-\delta_{IJ}$.

Our fields $\Psi_{+,I}^p$, $G_{+,I}$, and $w$ depend on $x^+$ and $y^I$.
The light-front formulation suggests to consider $x^+$ as dynamical time, 
and the 
transverse space $T$ as the space over which the initial configurations 
of the fields are set.

Let us write down $S_{glue}$ using the above Minkowski metric and the new field
variables. In terms of $E_{+}\equiv F_{-+}$, $E_I\equiv F_{-I}$, 
$F_{+I}$, and $F_{IJ}$, $S_{glue}$ reads
\begin{equation}
\label{Sglue}
S_{glue} = \int dx^+dy\Big[\frac{1}{2}\langle E_+|E_+\rangle + 
\langle E_I|F_{+I}\rangle -
\frac{1}{4}\langle F_{IJ}|F_{IJ}\rangle\Big].
\end{equation}
The integral in $x^+$ runs over $R$, and the integral in $y$ runs over $T$
(recall, that the integral in $x^-$ over the circle and the trace 
are  hidden in the definition (\ref{scalarprod}) of the scalar product). 

Use the complete orthonormal base $\chi_p$ to replace the scalar products
in (\ref{Sglue}) with the sums of the Fourier modes, separate the contribution
of the zero modes, take into account that the negative modes equal the
complex conjugation of the positive modes, and obtain
\begin{eqnarray}
\label{Smodes}
S_{glue}& =& \int dx^+dy\Big[\frac{1}{2} (E_+^0)^2 + 
E_I^0 F_{+I}^0 -
\frac{1}{4} (F_{IJ}^0)^2 +
\nonumber\\
&&+\sum_{p>0}\big(|E_+^p|^2  + 
2{\rm Re} (E_I^{-p} F_{+I}^p) -
\frac{1}{2} |F_{IJ}^p|^2\big)
\Big].
\end{eqnarray}

The next stage is to use the formulas from the previous section to express 
this action in terms of the Fourier data. Here we want to discuss two 
general properties of the emerging theory.

If $x^+$ is considered as time, $\Psi_+^p$ and $G_+$ are nondynamical, 
because there is no time derivatives of those fields in the action.
$\Psi_+^p$ can be integrated out, because the action is quadratic in it, and
the the relevant quadratic form is nondegenerate. The action is linear in $G_+$,
and variation over it generates a Gauss law which suffices to express
the derivatives of $w$ in terms of $\Psi_i^p$. In this way, we arrive at 
a theory whose field content is $\Psi_i^p$, $G_i$, and the zero mode of
$w$ (which is the integral of $w$ over the transverse space $T$).
This is the theory to which we apply the condition of a given
positive value of the momentum component of the system along the 
$x^-$ direction.

It was pointed out in \cite{Franke}, that, without 
an ultraviolet regularization, the Gauss law
of the light-front formulation has no solutions if localized 
charges are present on the transverse plane. The variables introduced above
help to clarify the situation. There are solutions to the Gauss law, but, in
the presence of charges on the transverse plane, 
they involve discontinuities of the Wilson loop $w$.

Generally, there are two options in such a situation. One is to consider
an ultraviolet regularization replacing the transverse space $T$ with a 
lattice. Another option is to introduce in the continuum theory 
the locations of the singularities 
of $w$ on the transverse plane as dynamical variables. 

We point out that the transverse lattice was introduced in 
\cite{Bardeen} (see \cite{Dalley} for recent references). Our formulation
provides a new option for introducing the transverse lattice in a gauge 
invariant way. It is natural to consider the matter fields $\Psi_i^p$
as defined on the sites of the transverse lattice, and replace the fields
$G_i$ by the phases defined on the links. Further work is needed to
develop this light-front program.

The authors are grateful to V. T. Kim for helpful comments.
This work was supported in part by the U. S. Department of Energy, 
grant No. DE-FG02-87ER40371, the NATO Science Program, grant No. PST.CLG.976521,
and by the Russian Foundation for Basic Research,
grant No. 00-02-17432.

\end{document}